\documentclass[12pt]{article}
\usepackage[utf8]{inputenc}
\usepackage[linesnumbered,ruled,vlined]{algorithm2e}
\usepackage{calrsfs}
\usepackage{graphicx,psfrag,epsf}
\usepackage{enumerate}
\usepackage{natbib}
\usepackage{subcaption}
\usepackage{booktabs}
\usepackage{epsfig}
\DeclareMathAlphabet{\pazocal}{OMS}{zplm}{m}{n}
\usepackage{amsmath,amsfonts,amsthm,bm} 
\usepackage{setspace}
\newcommand{\blind}{0}

\begin{document}

\def\spacingset#1{\renewcommand{\baselinestretch}%
{#1}\small\normalsize} \spacingset{1}


\if0\blind
{
  \title{\bf A Comparative Study of Imputation Methods for Multivariate Ordinal Data}
  \author{Chayut Wongkamthong \hspace{.2cm}\\
    Social Science Research Institute\\
    Duke University, USA\\
    chayut.wongkamthong@duke.edu\\
    \\
    and \\
    \\
    Olanrewaju Akande \\
    Social Science Research Institute, and\\
    The Department of Statistical Science\\
    Duke University, USA\\
    olanrewaju.akande@duke.edu}
  \maketitle
} \fi

\if1\blind
{
  \bigskip
  \bigskip
  \bigskip
  \begin{center}
    {\LARGE\bf A Comparative Study of Imputation Methods for Multivariate Ordinal Data}
\end{center}
  \medskip
} \fi

\bigskip
\begin{abstract}
Missing data remains a very common problem in large datasets, including survey and census data containing many ordinal responses, such as political polls and opinion surveys. Multiple imputation (MI) is usually the go-to approach for analyzing such incomplete datasets, and there are indeed several implementations of MI, including methods using generalized linear models, tree-based models, and Bayesian non-parametric models. However, there is limited research on the statistical performance of these methods for multivariate ordinal data. In this article, we perform an empirical evaluation of several MI methods, including MI by chained equations (MICE) using multinomial logistic regression models, MICE using proportional odds logistic regression models, MICE using classification and regression trees, MICE using random forest, MI using Dirichlet process (DP) mixtures of products of multinomial distributions, and MI using DP mixtures of multivariate normal distributions. We evaluate the methods using simulation studies based on ordinal variables selected from the 2018 American Community Survey (ACS). Under our simulation settings, the results suggest that MI using proportional odds logistic regression models, classification and regression trees and DP mixtures of multinomial distributions generally outperform the other methods. In certain settings, MI using multinomial logistic regression models is able to achieve comparable performance, depending on the missing data mechanism and amount of missing data. (total word counts: 6598 words)
\end{abstract}

\noindent%

\noindent
\textbf{Statement of significance}: This paper evaluates the repeated sampling properties of several multiple imputation (MI) methods for handling missing values in multivariate ordinal data. The methods include MI by chained equations (MICE) using multinomial logistic regression models, MICE using proportional odds models, MICE using classification and regression trees, MICE using random forest, MI using Dirichlet process (DP) mixtures of products of multinomial distributions, and MI using DP mixtures of multivariate normal distributions. The paper presents results from extensive simulations based on ordinal variables from real survey data --- the 2018 American Community Survey (ACS) --- and under different scenarios, including varying proportions of missing data and different missing data mechanisms. The results highlight several important conclusions about the performances of MI methods for ordinal data as well as several implementation decisions analysts will need to make in applying the methods. Finally, the results also illustrate some potential issues with uncertainty quantification when using some of the MI methods.
\\ 
\newline
This is a pre-copyedited, author-produced version of an article accepted for publication in Journal of Survey Statistics and Methodology following peer review. \newline
The version of record "Chayut Wongkamthong, Olanrewaju Akande, A Comparative Study of Imputation Methods for Multivariate Ordinal Data, Journal of Survey Statistics and Methodology, Volume 11, Issue 1, February 2023, Pages 189–212" is available online at: 
\url{https://doi.org/10.1093/jssam/smab028}.

\doublespacing
\section{Introduction} \label{intro}
Even as analysts and data scientists continue to gain access to larger and more complex datasets than ever before, in most fields, missing data remains a crucial problem. In large surveys for example, data is usually missing because participants often do not provide answers to some questions. 
It is well known that ignoring missing data can lead to biased inference and a reduction in efficiency, particularly when the observed and missing data are systematically different \citep{rubin:1976,littlerubin}. Analysts must therefore handle the missing data problem before or while analyzing such incomplete datasets.

The most common approach for handling missing data is multiple imputation (MI, \citet{rubin:1987}). In MI, one creates multiple completed versions of the sample dataset by replacing the missing entries with plausible draws from predictive distributions of the variables with missing values. These predictions are usually generated either through (i) joint modeling (JM), where one first specifies a multivariate distribution for all the variables and then predicts the missing values using the implied conditional distributions, or (ii) a fully conditional specification (FCS, \citet{VanBuuren2006, VanBuuren2007}), where one specifies a univariate model for each incomplete variable conditional on other variables, and uses the estimated univariate models to predict the missing values, without first forming a proper joint distribution. 

Several authors have developed several imputation engines under the JM approach \citep{Schafer:1997,sas_MI,amelia}.
Bayesian nonparametric models in particular, including Dirichlet process (DP) mixture models, have shown great promise as flexible imputation engines. When dealing with multivariate categorical data for example, one can use the DP mixture of products of multinomial distributions (DPMPM, \citet{si:reiter:13, manriquereiterjcgs, manrique:reiter:sm}) model. When dealing specifically with multivariate ordinal data, one can also use the DP mixture of multivariate normal distributions (DPMMVN, \citet{DPMMVN}) model, which uses latent variables with cut-offs to represent the ordinal variables. The DPMMVN model has an added advantage of being able to account for continuous covariates, when needed.

Many authors have also developed several imputation engines under FCS, arguably more than under JM, because of its simplicity and flexibility, even though it lacks the theoretical foundation of JM. Many of the implementations also often refer to the methodology as one of several names, including variable-by-variable imputation \citep{Brand1999}, sequential regression multivariate imputation (SRMI, \citet{Raghunathan2001}), multiple imputation by chained equations (MICE, \citet{VanBuurenOudshoorn2000, MICE}), and of course FCS \citep{VanBuuren2006, VanBuuren2007}, just to name a few. MICE in particular has perhaps become the most popular name for the method, through its software implementation. In this article, we also will henceforth refer to the method as MICE, and later write its software implementation in lower case as \textsf{mice}.

MICE is usually implemented by specifying generalized linear models (GLMs) for the univariate conditional (predictive) distributions \citep{Raghunathan2001}. For ordinal data, the most common GLMs used for MICE are multinomial or polytomous logistic regression models, and proportional odds or cumulative logistic regression models. For high dimensional datasets, one major drawback of MICE using GLMs is that including many interaction effects in each univariate model can be tedious and computationally intensive \citep{white2011}.
Thus, more recent implementations of MICE instead use machine learning and tree-based models as the conditional distributions. Two of the most popular tree-based methods for imputing multivariate categorical data, are classification and regression trees (CART, \citet{breiman:1984, burgreit10, Doove:2014}) and random forests \citep{breiman:2001,Shah:2014}. More recent imputation methods such as Generative Adversarial Imputation Nets (GAIN, \citet{GAIN}), use more advanced machine learning models to potentially improve upon some of the drawbacks of the other methods, including computational costs. GAIN for example, uses generative adversarial networks (GAN) to create a framework for missing data imputation. GAIN has been shown to do very well, in terms of overall predictive performance, in several machine learning datasets. In our evaluations however, we found the performance of GAIN to be inferior to the other imputation methods and thus only include the results in the supplementary material for reference.

There have been many empirical comparisons of several MI methods for multivariate continuous data, multivariate nominal data, or both \citep{VanBuuren2007, Diazordaz:2014, Laaksonen:2016, DPMPMcomparison_cat}. However, there are no rigorous empirical comparisons of MI methods for multivariate ordinal data. Specifically, to the best of our knowledge, there are no studies that provide practical guidance on potential advantages and disadvantages of using nominal models over the ordinal versions, when dealing with ordinal data.  Also, recent studies comparing the performance of newer machine learning imputation methods to more traditional methods like MICE \citep{missForest, Shah:2014,Bertsimas:2018, GAIN} have often focused on metrics for overall predictive power, for example, normalized root mean squared error (NRMSE) for continuous missing values, and proportion of falsely classified entries (PFC) for categorical missing values \citep{missForest, GAIN, deepGenerativeModel}. In fact, most have prioritized evaluating overall accuracy of the predictions, without extensive considerations of how the models preserve relationships in the data and properly account for uncertainty. Drawing conclusions on such empirical comparisons without fully accounting for the uncertainties, can be biased and misleading \citep{Shah:2014, wang2021deep}.

In this article, we compare six of the most popular MI methods for multivariate ordinal data, including (i) MICE using multinomial logistic regression models (MI-Multireg), (ii) MICE using proportional odds logistic regression models (MI-Polr), (iii) MICE using CART (MI-Cart), (iv) MICE using random forests (MI-Forest), (v) MI using DP mixtures of products of multinomial distributions (MI-DPMPM), and (vi) MI using DP mixtures of multivariate normal distributions (MI-DPMMVN). We compare all methods using evaluation metrics that assess how each method preserves and recovers both marginal and joint relationships in the data. We base our empirical comparisons on hypothetical populations comprising data from the 2018 American Community Survey (ACS).

The remainder of this article is organized as follows. In section \ref{Multiple Imputation for Ordinal Data}, we provide a review of the different imputation methods we assess in this study, and some implementation considerations. In section \ref{empirical_study}, we describe the ACS data, and the framework for our simulation study. We also define the evaluation metrics used to assess the quality of the imputed datasets under the methods. In section \ref{result}, we present the results of the simulation studies and discuss the performance of the different methods. In section \ref{conclusion}, we conclude with general takeaways.

\section{MI for ordinal data} \label{Multiple Imputation for Ordinal Data}

We first present the notation for describing MI for multivariate ordinal data.
Let $\bm{Y} = (\bm{Y}_1,\ldots,\bm{Y}_n)$ denote the $n \times p$ matrix containing the data for all $n$ observations across $p$ variables. For $i = 1,\ldots,n$, let $\bm{Y}_i = (Y_{i1},\ldots,Y_{ip})$, where each $Y_{ij} \in \{1,2,\ldots,D_j\}$ represents the value of variable $j$ out of $D_j$ levels for individual $i$ with $j = 1,\ldots,p$. Also, let $\bm{Y}_j = (Y_{1j},\ldots,Y_{nj})$. Let $\bm{R}$ denote the missing value indicator matrix where, for observation $i$ and variable $j$, $R_{ij} = 1$ when $Y_{ij}$ is missing, and $R_{ij} = 0$ otherwise. We partition $\bm{Y}_j$ as $\bm{Y}_j = (\bm{Y}_{j}^{obs},\bm{Y}_{j}^{mis})$, where  $\bm{Y}_{j}^{obs}$ represents all observed values for variable $j$, that is, data values in the $\bm{Y}_j$ vector corresponding to $R_{ij} = 0$, and $\bm{Y}_{j}^{mis}$ represents all missing values for variable $j$, that is, data values in the $\bm{Y}_j$ vector corresponding to $R_{ij} = 1$. We write $\bm{Y}^{obs} = (\bm{Y}_{1}^{obs},\ldots,\bm{Y}_{p}^{obs})$, and $\bm{Y}^{mis} = (\bm{Y}_{1}^{mis},\ldots,\bm{Y}_{p}^{mis})$. Finally, we write $\bm{Y} = (\bm{Y}^{obs}, \bm{Y}^{mis})$.

In MI, one creates $L > 1$ multiply-completed datasets $\textbf{Z} = (\textbf{Z}^{(1)}, \ldots, \textbf{Z}^{(L)})$. Each $\textbf{Z}^{(l)}$, with $l=1,\ldots,L$, is generated by replacing missing values in the sample $\bm{Y}$, with draws from the predictive distributions of models estimated based on the observed data $\bm{Y}^{obs}$. Analysts then can compute sample estimates for estimands of interest in each completed dataset $\textbf{Z}^{(l)}$, and combine them using MI inference rules developed by \citet{rubin:1987}. 

\subsection{MICE} \label{mice}

Under MICE, one imputes missing values by iteratively sampling plausible predicted values from a sequence of univariate conditional models, where the models are specified separately for each variable. First, one specifies an order for imputing the variables. The most popular choices are to (i) impute the variables in the order they appear in $\bm{Y}$, or (ii) impute the variables in increasing order of the number of missing cases. Next, one sets initial values for the missing entries. The most popular choice is to set the values by sampling from the marginal distribution of each $\bm{Y}_{j}^{obs}$. Alternatively, for each variable, one can also sample from its conditional distribution, given all other variables, where the distribution is constructed using only available cases.

After initialization, one cycles through the sequence of univariate conditional models, estimating and generating predictions from them, at each iteration. The process continues for $T$ total iterations until the chain converges \citep{abayomi}. We summarize the entire process in Algorithm \ref{MICE}.
\begin{algorithm}
\KwIn{Data matrix $\bm{Y}$ with columns sorted as per imputation order}
\KwOut{Multiple completed datasets $\textbf{Z}^{(1)}, \ldots, \textbf{Z}^{(L)}$}
\For{$l=1,\ldots,L$}{
Set $\bm{Y}^{(0)} = \bm{Y}$\\
Set initial values for missing entries in $\bm{Y}^{(0)}$\\
\For{$t=1,\ldots,T$}{
\For{$j=1,\ldots,p$}{
Fit the conditional model $(\bm{Y}_{j}^{obs} | \{\bm{Y}_k^{(t)}: k<j\}, \{\bm{Y}_k^{(t-1)}: k>j\})$\\
Generate $\bm{Y}_{j}^{mis(t)}$ from the implied $(\bm{Y}_{j}^{mis} | \bm{Y}_{j}^{obs}, \{\bm{Y}_k^{(t)}: k<j\}, \{\bm{Y}_k^{(t-1)}: k>j\})$
}
}
Set $\textbf{Z}^{(l)} = (\bm{Y}^{obs}, \bm{Y}^{mis(T)})$\\
}
\Return $\textbf{Z}^{(1)}, \ldots, \textbf{Z}^{(L)}$
\caption{MICE}
\label{MICE}
\end{algorithm}

\subsubsection{MI-Multireg and MI-Polr}
Most implementations of MICE use GLMs for the univariate conditional models in Step 6 of Algorithm \ref{MICE}. MI-Multireg and MI-Polr are the two most popular GLM models for implementing MICE for categorical data with more than two levels. MI-Multireg uses multinomial or polytomous logistic regression models \citep{Engel:1989}. Multinomial logistic regression models are the most common option for analyzing categorical data with more than two outcomes, making MI-Multireg a very popular MI method for both nominal and ordinal variables. MI-Polr on the other hand uses proportional odds logistic regression models \citep{McCullagh:1980, McCullagh:2005} for imputing missing data for ordinal variables. This results in a more parsimonious model than the multinomial logistic regression.
Even though MI-Polr may be expected to be more appropriate for ordinal data than MI-Multireg, we include the latter in the scope of our study to see whether there are certain settings where MI-Multireg would in fact outperform MI-Polr, when dealing with ordinal data.

To implement MI-Multireg, and MI-Polr, we use the ``polyreg'' and ``polr'' options respectively, in the \textsf{mice} package in \textsf{R} \citep{MICE}. We only consider main effects in the models and also keep most of the default options and arguments to mimic the default option most analysts often use. To generate $L=50$ multiple completed datasets for any given sample, under our simulation scenarios, using a standard notebook computer, MI-Multireg typically runs for approximately 1 hour and 30 minutes on average, and MI-Polr runs for about 50 minutes.

\subsubsection{MI-Cart} \label{MI-Cart}
In MI-Cart, one uses CART for the univariate conditional models in Step 6 of Algorithm \ref{MICE}, usually to provide more flexibility. CART for categorical data \citep{breiman:1984} is a nonparametric approach for modeling the relationship between a categorical response variable and potential predictors. CART follows a decision tree structure, using recursive binary splits to partition the predictor space into distinct non-overlapping regions. The top of the tree usually corresponds to the root and each successive split divides the space into two new branches further down the tree. The splitting or partitioning criterion at each node is chosen to minimize the ``information impurity'', for example, the Gini index of the two child nodes. Also, any splits that do not decrease the lack of fit by some certain thresholds are pruned off. The tree is then built until a stopping criterion, like the minimum number of observations in each child node, is met. 

When implementing MI-Cart, it is possible to use both regression trees and classification trees. Since classification trees are often more common and popular for categorical variables, we used them for imputation here.
Thus, to approximate the conditional distribution of any $\bm{Y}_j$, given a particular combination of the other variables, we use the proportion of values of $\bm{Y}_{j}^{obs}$ in the corresponding leaf. Specifically, to generate values for $\bm{Y}_{j}^{mis(t)}$ under MI-Cart, we traverse down the tree to the appropriate node using the combinations in $(\{\bm{Y}_k^{(t)}: k<j\}, \{\bm{Y}_k^{(t-1)}: k>j\})$, and sample from the $\bm{Y}_{j}^{obs}$ values in the leaf. For more details on MI-Cart, see \citet{burgreit10}.

We implement MI-Cart through the ``cart'' option in the \textsf{mice} package in \textsf{R}, and once again keep all default arguments, including the requirement that at least four observations must be in each of the terminal nodes. To generate $L=50$ multiple completed datasets for any given sample, under our simulation scenarios, using a standard notebook computer, MI-Cart typically runs for approximately 40 minutes.

\subsubsection{MI-Forest}
MI-Forest also uses a nonparametric approach for the univariate conditional models in Step 6 of Algorithm \ref{MICE}. MI-Forest uses random forests \citep{randomForest}, an ensemble tree method which builds multiple decision trees to the data. Specifically, random forest constructs many classification and regression trees using bootstrapped datasets, but only uses a sample of the predictors for each split in each tree. Doing so usually reduces the correlation among trees since it prevents the same variables from dominating the splitting process across all trees, and the decorrelation should result in predictions with less variance.

In MI-Forest, we again follow the general outline in Algorithm \ref{MICE}. That is, for each $\bm{Y}_j$ with missing data, we train a random forest model based on classification trees using available cases, given all other variables. We then generate plausible values for $\bm{Y}_{j}^{mis(t)}$ under that model. Next, we cycle through all the variables to generate one completed dataset. Finally, we run the entire process $L$ times to obtain $L$ imputed datasets.  For more details on MI-Forest, see \citet{Shah:2014}.

We implement MI-Forest through the ``rf'' option in the \textsf{mice} package in \textsf{R}, and once again keep all default arguments, including growing ten trees by default. To generate $L=50$ multiple completed datasets for any given sample, under our simulation scenarios, using a standard notebook computer, MI-Forest typically runs for approximately 40 minutes.

\subsection{MI-DPMPM}
MI-DPMPM uses the DP mixture of products of multinomial distributions (DPMPM) model, to generate imputations for multivariate categorical data containing missing values \citep{si:reiter:13}. The DPMPM model, which treats each variable as nominal, assumes that each observation $i$ in the data belongs to a latent class $z_i \in \{1,\ldots,K\}$, and variables within each latent class $k = 1,\ldots,K$, follow independent multinomial distributions. 

Specifically, let $\pi_k = P(z_i = k)$, for $k = 1,\ldots,K$, be the probability that observation $i$ belongs to latent class $k$. Let $\lambda_{kjd} = P(Y_{ij} = d|z_i = k)$ be the probability that $Y_{ij}$ takes value $d$, given that it belongs to latent class $k$. Finally, let $\bm{\pi} = (\pi_1,\ldots,\pi_K)$ and $\bm{\lambda} = \{\lambda_{kjd}: k = 1,\ldots,K; j = 1,\ldots,p; d = 1,\ldots,D_j\}$. We write the generative model as
\begin{align}
Y_{ij}|z_i, \bm{\lambda} &\stackrel{ind}{\sim} \text{Discrete}(\lambda_{z_ij1},\ldots,\lambda_{z_ijD_j}) \quad \text{ for all $i$ and $j$} \label{DPMPM:level1}\\ 
z_i| \bm{\pi} &\stackrel{iid}{\sim} \text{Discrete}(\pi_1,\ldots,\pi_K) \quad \text{ for all $i$}. \label{DPMPM:level2}
\end{align}
Under this model specification, averaging over the $K$ latent classes induces marginal dependence between the variables. With a large enough number of classes $K$, the model is consistent for any joint probability distribution \citep{DPMPM_Kenough}.

For prior distributions, we follow \citet{schifeling:reiter, si:reiter:13, manrique:reiter:sm}. We use independent uniform Dirichlet distributions for each probability vector in $\bm{\lambda}$, and the truncated stick breaking representation of the DP for the mixture probabilities. We have
\begin{align}
(\lambda_{kj1},\ldots,\lambda_{kjD_j})  &\stackrel{ind}{\sim} \text{Dirichlet}(\bm{1}_{D_j}) \quad \text{ for all $j$ and $k$}\\
\pi_k & = V_k\prod_{h<k}(1-V_h) \label{stick-break-1}\\
V_k | \alpha &\stackrel{iid}{\sim} \text{Beta}(1,\alpha), \quad \text{for } k=1,\ldots,K-1; \quad V_K = 1 \label{stick-break-2}\\
\alpha &\sim \text{Gamma}(0.25, 0.25). \label{stick-break-3}
\end{align}
We obtain posterior samples of all parameters in the model using Gibbs sampling.  We also follow \citet{si:reiter:13} when setting $K$, starting with a relatively modest value during initial runs. Whenever the number of occupied classes reaches $K$ across any of the iterations, we gradually increase $K$ until that is no longer the case. Finally, we handle missing values directly within the Gibbs sampler. At any iteration $t$, we sample each $Y_{ij}^{mis(t)}$ value in $\bm{Y}_{j}^{mis(t)}$ from \eqref{DPMPM:level1}, conditional on the current draw of the parameters, and the latent class $z_i$. To obtain $L$ multiple imputed datasets, we select $L$ evenly spaced out draws from the posterior predictive samples over the iterations, after the chain converges. For more details on the DPMPM model, including further discussions on prior specifications and setting $K$, see \citet{si:reiter:13, manrique:reiter:sm}.

We use the \textsf{NPBayesImputeCat} package in \textsf{R} \citep{manriquereiterjcgs} to implement MI-DPMPM. We set the number of latent classes $K=40$, based on initial runs on sample datasets of size 10,000. For each simulation, we run the Gibbs sampler for 15,000 iterations, and set the first 5,000 as burn-in. We assess the convergence of the sampler by examining trace plots of combinations of the model parameters, such as random samples of marginal probabilities, that are insensitive to label switching. To generate $L=50$ multiple completed datasets for any given sample, under our simulation scenarios, using a standard notebook computer, MI-DPMPM typically runs for approximately 5 minutes.

\subsection{MI-DPMMVN} \label{DPMMVNsection}
MI-DPMMVN uses the DP mixture of multivariate normal distributions (DPMMVN) model \citep{DPMMVN}, to generate imputations for multivariate ordinal data.  Like the DPMPM, the DPMMVN model also assumes that each observation $i$ in the data belongs to a latent class $z_i \in \{1,\ldots,K\}$. However, unlike the DPMPM, the DPMMVN model is primarily designed for modeling ordinal response variables. The DPMMVN model assumes that the ordinal variables arise from continuous latent variables. The latent variables (as well as continuous predictors when available) are assumed to jointly follow a multivariate normal distribution conditional on the latent classes.

Specifically, suppose each $Y_{ij} \in \{1,2,\ldots,D_j\}$ as before. We introduce continuous latent random variables $\bm{x}_i = (x_{i1}, \ldots,x_{ip})$ for all $i$, and cut-offs $\bm{\gamma_j} = (\gamma_{j0},\ldots,\gamma_{jD_j})$ for each $j$. Each $Y_{ij} = d$ if and only if the corresponding $x_{ij} \in (\gamma_{j(d-1)}, \gamma_{jd}]$, where $d = 1,2,\ldots,D_j$. These latent variables follow a DP mixture of multivariate normal distributions. We have
\begin{align}
\bm{x}_i|z_i,\bm{\mu}_{z_i}, \bm{\Sigma}_{z_i} &\stackrel{ind}{\sim} N(\bm{\mu}_{z_i}, \bm{\Sigma}_{z_i}) \quad \text{ for all $i$ and $j$} \label{DPMMVN:level1}\\
z_i| \bm{\pi} &\stackrel{iid}{\sim} \text{Discrete}(\pi_1,\ldots,\pi_K) \quad \text{ for all $i$},
\end{align}
combined with the truncated stick breaking DP prior in \eqref{stick-break-1} to \eqref{stick-break-3}. For prior distributions on the remaining parameters, we follow \citet{DPMMVN} and specify
\begin{align}
\bm{\mu}_{k}, \bm{\Sigma}_{k} |  \bm{m}, \bm{V}, \bm{S} &\stackrel{iid}{\sim} N(\bm{m}, \bm{V}) \cdot IW(\nu, \bm{S}) \quad \text{ for all $k = 1,\ldots,K$}\\
\bm{m} & \sim N(\bm{a}_m, \bm{B}_m); \quad \bm{V} \sim IW(a_V, \bm{B}_V); \quad \bm{S} \sim W(a_S, \bm{B}_S) \label{hyperprior1}.
\end{align}
Here, $W(a_S, \bm{B}_S)$ is a Wishart distribution with mean $a_S\bm{B}_S$, and $IW(a_V, \bm{B}_V)$ is an inverse-Wishart distribution with mean $(a_v-p-1)^{-1}\bm{B}_V$. Also, $\bm{m}$ represents the overall mean across all the groups (mean of the group-specific means), $\bm{V}$ characterizes how the group-specific means vary around $\bm{m}$, and $\bm{S}$ represents the overall covariance across all the groups (scaled mean of the group-specific covariances). We set the parameters in \eqref{stick-break-1} to \eqref{hyperprior1} to represent vague prior specifications following recommendations by \citet{DPMMVN}. To set $K$, we can use a similar strategy as in DPMPM. The cut-offs for each variable can be fixed beforehand at arbitrary values and the covariance matrices will be identifiable \citep{DPMMVN}.

We again use Gibbs sampling to generate posterior draws of the parameters, and handle missing values directly within the sampler, just as we did with the DPMPM. Specifically, for each missing entry, we sample the latent variables using \eqref{DPMMVN:level1}, and use the cut-offs to impute the corresponding ordinal level. For more details, see \citet{DPMMVN}.

To implement MI-DPMMVN, we adapt the \textsf{R} code in \citet{DPMMVN}'s supplementary material. We set the number of latent classes $K=50$ based on initial runs. Similar to MI-DPMPM, we run the Gibbs sampler for each simulation run for 15,000 MCMC iterations, and set the first 5,000 as burn-in. Again, we assess convergence using parameters and estimands that are insensitive to label switching. To generate $L=50$ multiple completed datasets for any given sample, under our simulation scenarios, using a standard notebook computer, MI-DPMMVN typically runs for about 1 hour and 30 minutes.

\subsection{Implementation considerations}

In applying the MI methods discussed above, analysts will need to make several decisions. First, analysts must make a decision on whether to use MICE or JM based methods. MICE is often preferred in practice because of its simplicity and flexibility in specifying different models for the conditional distributions. However, a major drawback is that the conditional distributions themselves may be incompatible, in the sense that they do not correspond to a joint distribution. This can make JM-based methods more desirable. On the other hand, specifying accurate joint distributions for a large number of variables in practice can be very challenging, especially when dealing with different variable types. Thus, many JM approaches, including for example those implemented in \textsf{proc MI} in \textsf{SAS} \citep{sas_MI}, \textsf{AMELIA} \citep{amelia}, and \textsf{norm} in \textsf{R} \citep{Schafer:1997}, often make a simplifying assumption that the data follow multivariate normal distributions, even for categorical data, and usually with varying levels of success depending on the settings \citep{KropkoEtAl2014, KarangwaEtAl2016}. Other JM methods such as MI-DPMPM and MI-DPMMVN do not make this simplifying assumption, however, they can be very computationally expensive. Even with the theoretical drawback however, MICE has been shown to work remarkably well in practice and outperform many theoretically sound JM-based methods when dealing with many variable types \citep{van2018flexible}. We see similar results for some of the MICE methods in our simulations. We therefore recommend that analysts take the different advantages and disadvantages of both MICE and JM into consideration when selecting their imputation methods.

Second, analysts must make a decision about whether or not to use parametric models under either JM or MICE. Overall, MI implementations based on nonparametric models or machine learning models, for example, MI-Cart and MI-Forest, usually can provide flexibility in model specification especially over those based on GLMs, since the former (i) do not rely heavily on parametric assumptions of the latter, (ii) do not require the tedious and intensive process of selecting many interaction effects in high dimensional datasets, as is often the case with the latter, and (iii) can usually handle mixed variable types more easily than the latter. However, the results of the MI implementations based on machine learning models in particular can vary substantially depending on the level of tuning required. In MI-Forest for example, tuning parameters such as the number of trees and the number of nodes often need to be investigated carefully. Other more complex methods such as GAIN \citep{GAIN} -- though once again not a major focus of our study -- can require a more substantial level of tuning. Thus, when the model assumptions are likely to hold, it may be more beneficial to rely on MI implementations based on GLMs. Again, we recommend that analysts investigate the assumptions of each method carefully before using them.

Finally, analysts will need to make a decision on the software implementations to use for the methods, as poor implementations of any of the methods can have significant implications on the performance of the method. For example, in the supplementary material, we compare the results of two implementations of MI-Forest, one through the \textsf{mice} package in \textsf{R} and the other through the \textsf{missForest} package in \textsf{R}. The \textsf{missForest} package always imputes predicted values based on majority rule, but the \textsf{mice} package on the other hand ensures that the imputed values are actually approximate random draws from the predictive distribution. Thus, the former usually introduces bias in the imputed data and the results in the supplementary material reflect this. Similar issues may also exist for the different software implementations of the other methods. In general, we recommend that analysts first investigate the differences and potential limitations of any implementation of each imputation method before using them. Consequently, we note that our results and discussions in this paper are based on the specific software implementations of the imputation methods we have chosen and may not extend to other implementations.

\section{Empirical study} \label{empirical_study}

\subsection{Framework and ACS data}

We conduct our empirical comparison of the methods using data from the 2018 American Community Survey (ACS) Public Use Microdata Sample (PUMS) files. As is generally the case for single-year ACS PUMS, the 2018 ACS PUMS contains sample data from about 1 in every 40 households in the United States. The data includes over 500 variables collected both at the individual level (e.g., income, age, sex, educational attainment) and at the household level (e.g., lot size, and number of rooms, vehicles, and persons). At the household level, the original dataset contains 1,385,396 records while we have 3,214,539 records at the individual level. We only focus on a subset of ordinal variables from this dataset. The data can be downloaded from the United States Bureau of the Census \textit{(https://www2.census.gov/programs-surveys/acs/data/pums/2018/}).

For this study, we combine housing unit records with the household head records from the individual level data, and treat this as our unit of analysis. We do so to create a richer set of ordinal variables than what is available at the household level alone. Also, we use individual level data for the household head alone because individuals are nested within households by design, which violates the independence assumption needed for the MI methods.
We remove all empty housing units and identification variables, including serial number, state, area code, and division. This results in data containing 1,257,501 records and 11 ordinal variables, which we treat as our population data, from which we repeatedly sample from. The 11 variables comprise of 1 variable with two levels, 3 variables with three to five levels, 4 variables with six to ten levels, and 3 variables with more than ten levels. We describe the variables in more detail in the supplementary material.

We follow the approach in \citet{DPMPMcomparison_cat} and use repeated sampling to evaluate the statistical properties of the imputation methods. Here, we randomly sample 10,000 units from the population, without replacement, and create varying proportions of missing data, under both the missing completely at random (MCAR, \citet{littlerubin}) and missing at random (MAR, \citet{littlerubin}) scenarios. First, we set five of the variables to be fully observed. We do so to mimic real applications where some of the variables are often fully observed. Next, we set either 30\% or 45\% of the values of the remaining six variables to be missing according to either MCAR or MAR mechanisms. Across all datasets, the missing data scenarios tend to result in complete cases ranging from 1\% (for 45\% MAR) to 12\% (for 30\% MAR), making complete case analysis untenable.
\begin{figure}[t!]
\centerline{\includegraphics[scale=0.75]{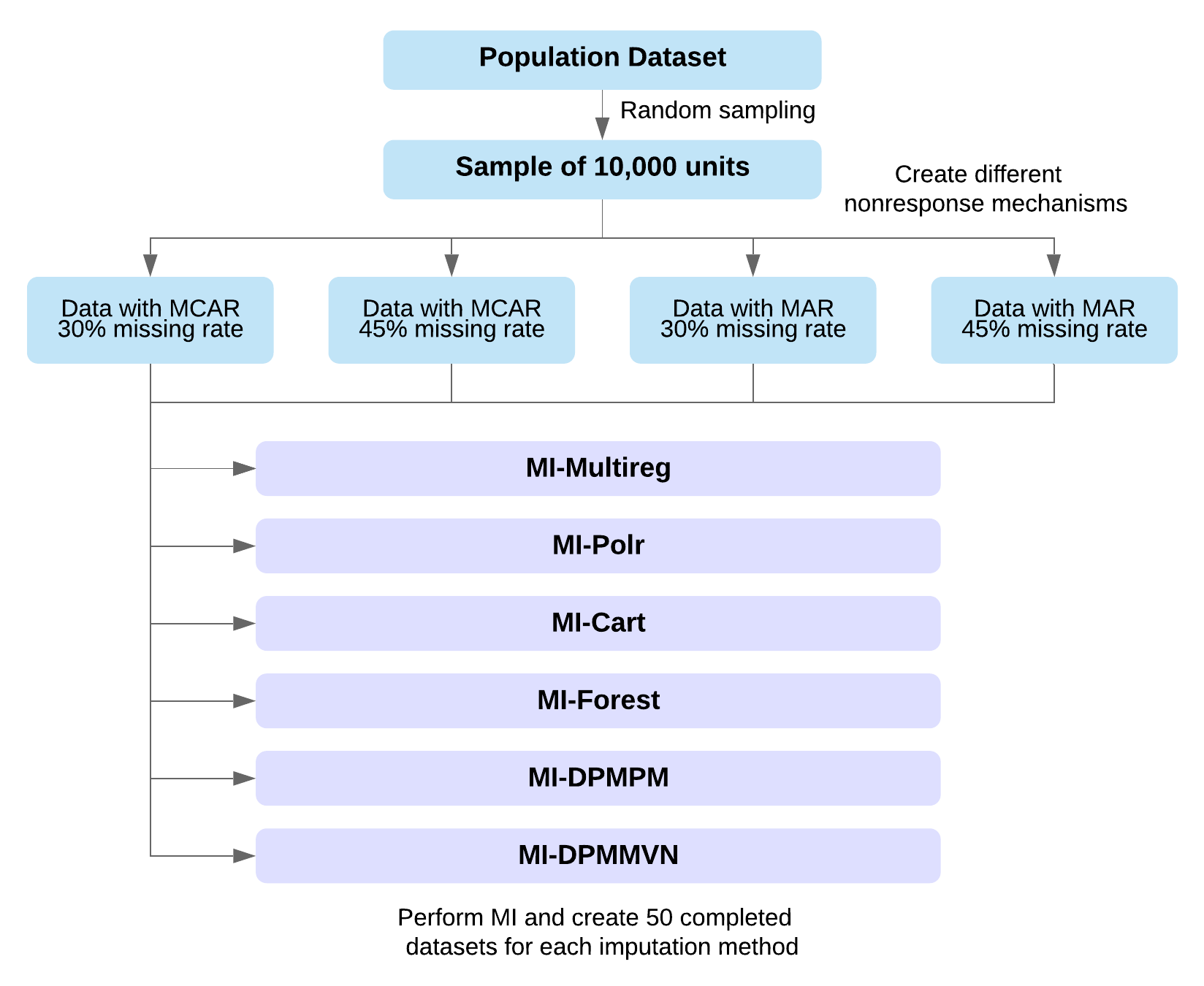}}
\caption{Imputation process for a single random sample from the population.}
\label{experiment}
\end{figure}
For each random sample of 10,000 observations, and under each combination of missingness mechanism and proportion of missing data, we use each imputation method to create $L = 50$ multiple completed datasets. Figure \ref{experiment} describes the process for each random sample of 10,000 observations. We repeat the entire process 500 times, each time generating new samples from the population and new missingness patterns.

\subsection{Performance metrics} \label{metrics}
We compare the performances of the MI methods using coverage rates, relative MSE, and bias, based on the multiply-imputed datasets. We do so to evaluate accuracy in estimating joint relationships within the data, rather than using one-number metrics that only probe overall accuracy. In practice, the end-goal of doing MI is usually parameter estimation in an analysis model or summary statistics of the variables. Here, we focus on the latter and thus evaluate the performance metrics on sets of marginal probabilities, bivariate probabilities, and trivariate probabilities separately. We compute estimates of these probabilities in each completed dataset and combine them using \citet{rubin:1987}'s combining rules. We present a quick review of the rules in the supplementary material.


Let $q$ be the completed-data point estimator of estimand $Q$. For $l=1, \dots, L$, let $q_l$  be the values of $q$ in completed dataset $\textbf{Z}^{(l)}$.  Then, the MI point estimate of $Q$ is $\bar{q}_L = \sum_{l=1}^{L}q_l/L$. MI inferences require that the central limit theorem applies to the complete-data estimate and that the sampling distribution of the complete-data estimate is approximately normal. In our simulations, we treat the value of $Q$ in the population data as the ``ground truth''. We therefore follow \citet{DPMPMcomparison_cat} and only consider estimands that satisfy $nQ > 10$ and $n(1-Q) > 10$, to eliminate estimands where the central limit theorem is not likely to hold even with no missing data. 

To compute coverage rates, we first construct 95\% confidence intervals for each estimand, using the MI inference rules. We then compute the proportion of the five hundred 95\% confidence intervals that contain the corresponding Q from the full population data. High
quality of the imputations would generally imply coverage rates close to the nominal rate of 0.95 or close to the corresponding coverage rates in the ``pre-missing data''. For relative MSE, we compare the MSE of each estimator from the imputed datasets, to the MSE from the original sample without missing data. That is,
\begin{align}
\text{Relative MSE} & = \frac{\text{MSE based on imputed data}}{\text{MSE based on pre-missing data}} = \frac{\sum_{h=1}^{500}[\bar{q}^{(h)}_L-Q]^2}{\sum_{h=1}^{500}[\hat{q}^{(h)}-Q]^2}. \label{relMSE}
\end{align}
Here, $\bar{q}^{(h)}_L$ is the value of $\bar{q}_L$ from sampled dataset $h$ while $\hat{q}^{(h)}$ is the point estimate of $Q$ from the same dataset before introducing missing values. Lower values of relative MSE, especially values closer to one, would indicate higher quality of imputations. 
Finally, we compute the estimated bias under each method. This allows us to examine whether the differences in the relative MSE values are due to the estimated bias values.
We have
\begin{align}
\text{Bias} & = \left[\frac{1}{500}\sum_{h=1}^{500}\bar{q}_L^{(h)}\right] - Q.
\end{align}
Lower values of bias would clearly indicate higher quality of imputations.

\section{Simulation results} \label{result}
We summarize the results for all the  MI methods using graphical displays and tables. We present the most important findings for the MAR and MCAR scenarios, with missingness rates of 30\% and 45\% for the six variables with missing data. We also focus primarily on results for marginal and bivariate probabilities. We present additional results and findings, including the results for trivariate probabilities, in the supplementary material. 

\subsection{MAR scenarios}
We begin with an evaluation of all the MI methods under MAR, since it is often the most plausible missing data mechanism in practice. First, we create a simulation scenario with $n = 10,000$ and $30\%$ missingness rate. To incorporate MAR, we set missing values independently for two variables (number of vehicles available and period since the household head last worked), and then generate missingness indicators, for four other variables ( the number of persons in household, educational attainment, age of household head, and total person's income), from logistic regression models, conditional on the remaining five fully observed variables. For simplicity, we only include main effects in the logistic regression models, and we tune the coefficients to result in approximately $30\%$ missing data, per variable. We present the specified logistic regression models in the supplementary material.

\begin{figure}[t!]
\centerline{\includegraphics[scale=.54]{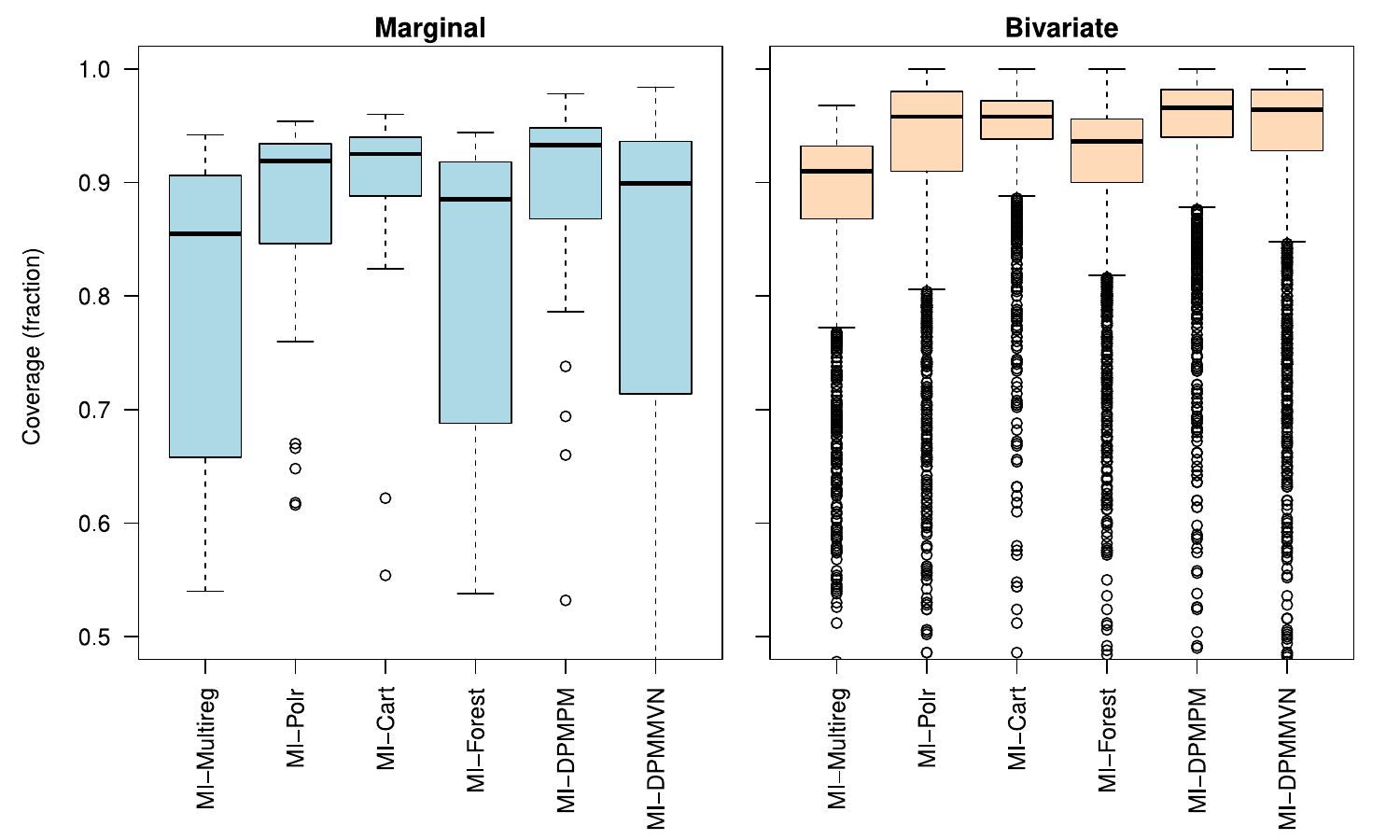}}
\caption{Distributions of coverage rates for all MI methods, under the 30\% MAR scenario. We set the lower limit of the y-axis scale to 0.5 to zoom in on the results. We present numerical summaries of the entire distributions in the supplementary material. }
\label{coverage_plot_MAR_30}
\end{figure}
Figure \ref{coverage_plot_MAR_30} displays the estimated coverage rates of the $95\%$ confidence intervals for the marginal and bivariate probabilities, under each method. The distributions of coverage rates for most of the methods are visibly left skewed, especially for the bivariate probabilities. MI-DPMMVN has the most skew of all the methods. The extremely low coverage rates across all the methods often correspond to combinations of the variables with very low probabilities in the full population. MI-Polr, MI-Cart, MI-DPMPM, all result in coverage rates that are generally closest to the nominal $95\%$ level, with comparable tails. In fact, the median coverage rates for all three methods are above $0.90$ for marginal and bivariate probabilities. The coverage rates for MI-Cart in particular tend to have the least skew overall. MI-Multireg tends to result in median coverage rates that are farthest from the nominal $95\%$ level, for marginal and bivariate probabilities, although MI-Forest and MI-DPMMVN do not appear to be visibly better.

\begin{table}[t!]
\centering
\footnotesize
\caption{Distributions of relative MSEs for marginal and bivariate probabilities across all MI methods, under the 30\% MAR scenario.} 
\begin{tabular}{rrrrrrr}
  \toprule
& MI-Multireg & MI-Polr & MI-Cart & MI-Forest & MI-DPMPM & MI-DPMMVN \\ 
  \midrule
  \multicolumn{1}{c}{}  &   \multicolumn{6}{c}{Marginal} \\
Min. & 1.35 & 1.25 & 1.26 & 1.30 & 1.26 & 1.02 \\ 
  1st Qu. & 1.92 & 1.56 & 1.48 & 1.67 & 1.59 & 1.54 \\ 
  Median & 2.56 & 1.84 & 1.71 & 2.20 & 1.94 & 2.16 \\ 
  3rd Qu. & 6.80 & 2.32 & 2.35 & 3.63 & 2.83 & 4.60 \\ 
  Max. & 56.74 & 22.74 & 8.56 & 18.25 & 8.87 & 234.35 \\ 
  \multicolumn{7}{c}{} \\
  \multicolumn{1}{c}{}  &   \multicolumn{6}{c}{Bivariate} \\
Min. & 0.99 & 0.46 & 0.36 & 0.45 & 0.41 & 0.56 \\ 
  1st Qu. & 1.42 & 0.96 & 1.05 & 1.12 & 0.96 & 0.95 \\ 
  Median & 1.96 & 1.18 & 1.25 & 1.42 & 1.13 & 1.12 \\ 
  3rd Qu. & 2.91 & 2.04 & 1.65 & 1.88 & 1.59 & 1.87 \\ 
  Max. & 200.52 & 231.68 & 25.30 & 30.19 & 37.68 & 204.85 \\ 
  \bottomrule
\end{tabular}
\label{relMSE_MAR_30}
\end{table}
Table \ref{relMSE_MAR_30} displays the distributions of the estimated relative MSE for the marginal and bivariate probabilities, under each method. For marginal probabilities, MI-Polr, MI-Cart, and MI-DPMPM tend to yield point estimates with the lowest MSE, relative to the MSE of the pre-missing data. Overall, across both the marginal and bivariate estimands, MI-Cart and MI-DPMPM generally result in the smallest relative MSE values, when looking at the entire distributions of the relative MSE values, thus outperforming all other methods under this simulation setting. For bivariate probabilities, MI-Polr still produces comparable relative MSE values to MI-Cart and MI-DPMPM but has much longer tails for large values. The performance of MI-Forest is often not too far behind those three. MI-Multireg and MI-DPMMVN tend to result in the highest relative MSE values. We note that while MI-DPMMVN generally results in relative MSE values that are higher than MI-Polr for marginal estimands, the trend largely reverses when looking at bivariate estimands. Essentially, MI-DPMMVN outperforms both MI-Polr and MI-Multireg when estimating bivariate probabilities. This suggests that when looking at measures of accuracy such as MSE, MI-DPMMVN may actually be more likely to yield more accurate estimates for joint relationships, than the MI using GLM methods.

\begin{figure}[t!]
\centerline{\includegraphics[scale=0.55]{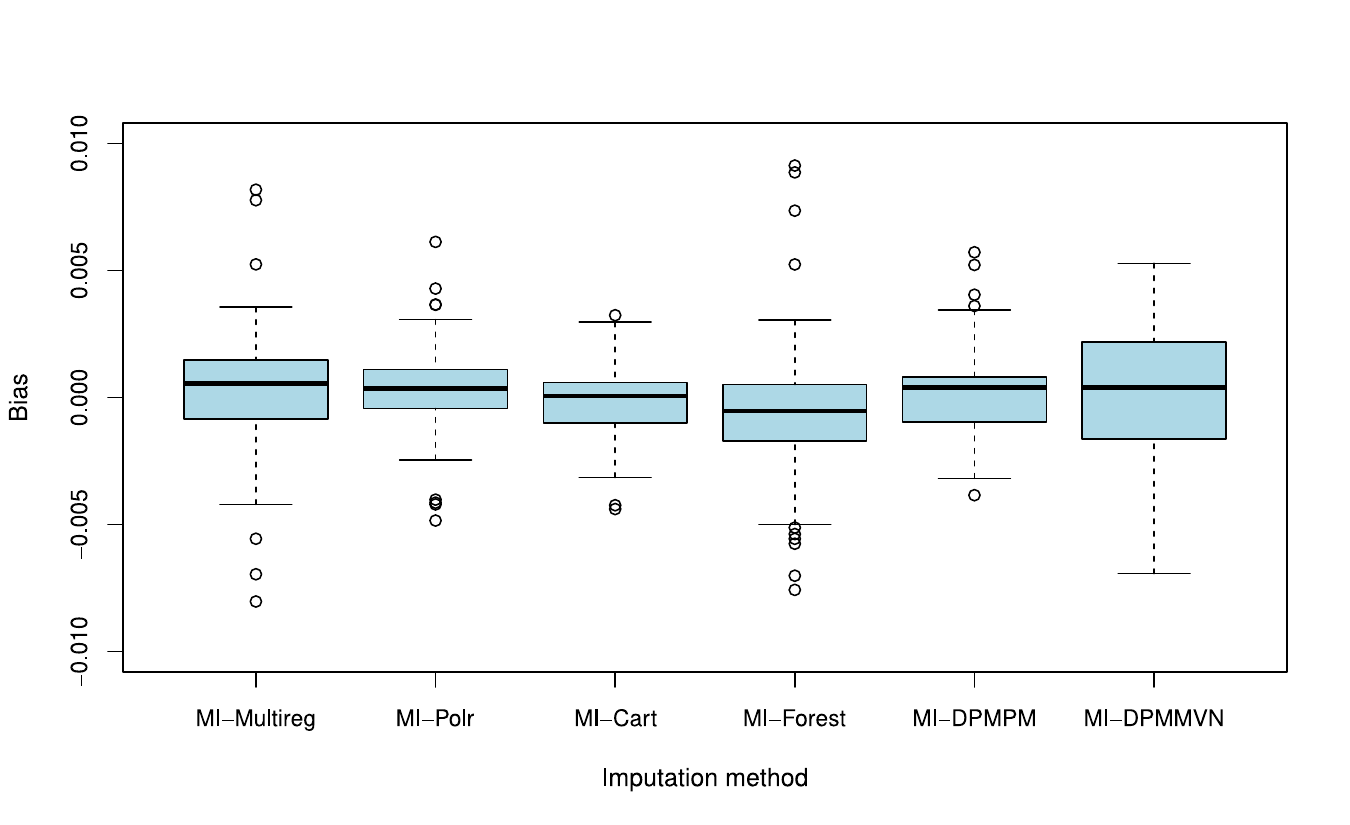}}
\caption{Distributions of estimated bias for marginal probabilities across all MI methods, under the 30\% MAR scenario.}
\label{bias_plot_MAR_30}
\end{figure}

To provide additional context for the trends in the relative MSE values, we also examine the estimated bias for all the methods. Figure \ref{bias_plot_MAR_30} displays the distributions of estimated bias for the marginal probabilities under each method.
The overall trends appear to mimic the trends in the estimated relative MSE, at least for marginal estimands. MI-Cart, MI-Polr, and MI-DPMPM, again appear to outperform all other methods, that is, their distributions of estimated bias are closer and more concentrated around zero than the other methods. We present numerical summaries for the figure and the estimated bias for bivariate and trivariate probabilities in the supplementary material.

\begin{table}[t!]
\centering
\footnotesize
\caption{Distributions of relative MSEs for marginal and bivariate probabilities across all MI methods, under the 45\% MAR scenario.}
\begin{tabular}{rrrrrrr}
  \toprule
 & MI-Multireg & MI-Polr & MI-Cart & MI-Forest & MI-DPMPM & MI-DPMMVN \\ 
  \midrule
  \multicolumn{1}{c}{}  &   \multicolumn{6}{c}{Marginal} \\
Min. & 1.69 & 1.55 & 1.57 & 1.54 & 1.56 & 1.11 \\ 
  1st Qu. & 2.90 & 1.96 & 1.93 & 2.48 & 2.30 & 2.30 \\ 
  Median & 5.37 & 2.41 & 2.44 & 3.05 & 2.76 & 3.66 \\ 
  3rd Qu. & 14.24 & 3.78 & 3.71 & 7.73 & 4.30 & 9.01 \\ 
  Max. & 284.10 & 1101.21 & 22.80 & 49.24 & 21.93 & 590.62 \\ 
  \multicolumn{7}{c}{} \\
  \multicolumn{1}{c}{}  &   \multicolumn{6}{c}{Bivariate} \\
Min. & 0.99 & 0.36 & 0.36 & 0.43 & 0.50 & 0.42 \\ 
  1st Qu. & 1.82 & 1.01 & 1.13 & 1.29 & 1.00 & 1.00 \\ 
  Median & 3.15 & 1.51 & 1.53 & 1.85 & 1.37 & 1.51 \\ 
  3rd Qu. & 5.51 & 3.46 & 2.40 & 2.98 & 2.32 & 3.14 \\ 
  Max. & 373.98 & 1575.66 & 27.98 & 81.57 & 61.79 & 522.97 \\ 
   \bottomrule
\end{tabular}
\label{relMSE_table_MAR_45}
\end{table}

As a sensitivity analysis, we examine whether the overall conclusions from our MAR scenario change as the missingness rate increases. We create the same MAR scenario as before but tune all parameters to result in $45\%$ missingness rate, for each of the six variables with missing data. Table \ref{relMSE_table_MAR_45} displays the distributions of the estimated relative MSE for the marginal and bivariate probabilities, under each method. As expected with a higher missingness rate, all relative MSE values are generally larger across all methods. For all the methods and for most of the estimands, the relative MSE values for the marginal estimands increase by a multiplicative factor of approximately 1.3 to 2, compared to the $30\%$ MAR scenario. The overall trends for the marginal and bivariate probabilities are very similar to the $30\%$ MAR scenario, with MI-Polr, MI-Cart, MI-Forest and MI-DPMPM, once again yielding the most accurate point estimates. Although, the largest differences in the relative MSE values, when comparing this scenario to the $30\%$ MAR scenario, correspond to MI-Multireg, MI-Polr actually has the largest maximum relative MSE values overall. Across all methods, the maximum values result from one of the four variables with missing values under MAR. Two levels of this variable have very low probabilities in the original data. With $45\%$ missing data, there is very little data in combinations involving that level, and interestingly, the proportional odds model appears to struggle the most from this.

We also present additional results for this scenario, including the estimated marginal, and joint probabilities for each MI method, as well as the distributions of estimated coverage rates, and bias, in the supplementary material. Qualitatively, the overall trends are the same when we increase the missingness rate from $30\%$ to $45\%$.

\subsection{MCAR scenarios}
\begin{figure}[t!]
\centerline{\includegraphics[scale=.54]{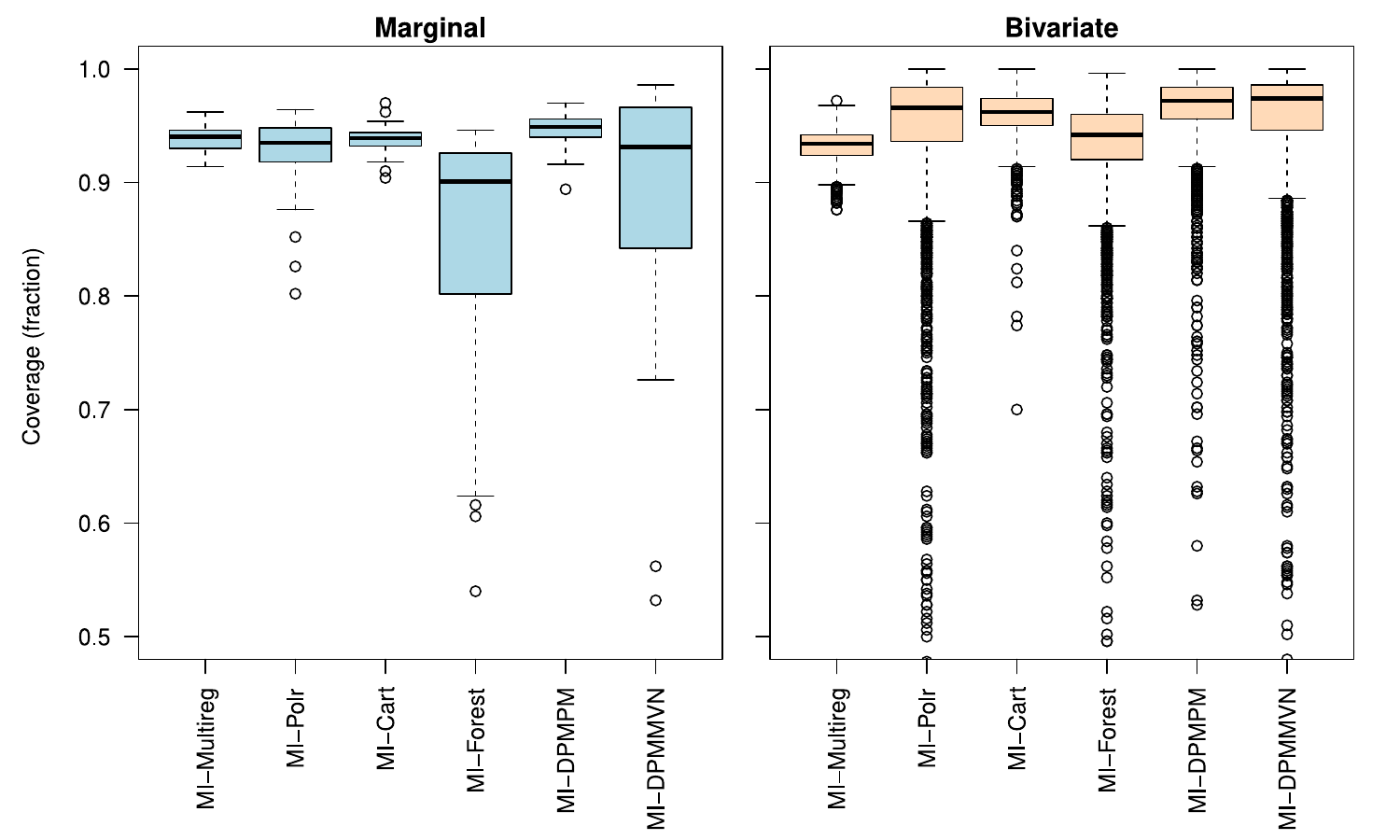}}
\caption{Distributions of coverage rates for all MI methods, under the 30\% MCAR scenario.}
\label{coverage_plot_MCAR_30}
\end{figure}

Analysts still sometimes have to conduct analyses under MCAR. Thus, we create a simulation scenario with $n = 10,000$ and $30\%$ missingness rate. To incorporate MCAR, we set missing values independently for the six variables. Figure \ref{coverage_plot_MCAR_30} displays the estimated coverage rates of the $95\%$ confidence intervals for the marginal and bivariate probabilities.
The coverage rates across all methods are generally higher than the coverage rates from the MAR scenarios. MI-Polr, MI-Cart, MI-DPMPM, once again yield more accurate point estimates than the rest. Interestingly, under this scenario, the median coverage rate (and overall distribution of the rates) for MI-Multireg is actually higher than the median coverage rate (and overall distribution of the rates) for MI-Polr for marginal estimands. For bivariate estimands, the trend reverses, with MI-Polr having the higher median coverage rate. However, MI-Polr has a much longer lower tail, sometimes reaching very low rates. This finding is also consistent with the result of the 45\% MCAR scenario in the supplementary material.

Table \ref{relMSE_table_1_MCAR_30} displays the distributions of the estimated relative MSEs for the marginal and bivariate probabilities, under each method.  
Again, MI-Cart and MI-DPMPM generally outperform all other methods, especially for bivariate probabilities, although not by much for most of the distributions. MI-Polr once again has the largest maximum relative MSE values across all methods, for the bivariate probabilities. The results also suggest that MI-Multireg can yield much more comparable results to MI-Polr in particular, when dealing with MCAR, rather than MAR. In fact, MI-Multireg has the smallest maximum relative MSE values across all methods, for the bivariate probabilities.
\begin{table}[t!]
\centering
\footnotesize
\caption{Distributions of relative MSEs for marginal and bivariate probabilities across all MI methods, under the 30\% MCAR scenario.} 
\begin{tabular}{rrrrrrr}
  \toprule
 & MI-Multireg & MI-Polr & MI-Cart & MI-Forest & MI-DPMPM & MI-DPMMVN \\ 
  \midrule
    \multicolumn{1}{c}{}  &   \multicolumn{6}{c}{Marginal} \\
Min. & 1.22 & 1.18 & 1.21 & 1.21 & 1.18 & 0.82 \\ 
  1st Qu. & 1.36 & 1.41 & 1.35 & 1.50 & 1.37 & 1.09 \\ 
  Median & 1.44 & 1.47 & 1.40 & 1.61 & 1.45 & 1.64 \\ 
  3rd Qu. & 1.52 & 1.67 & 1.44 & 2.47 & 1.52 & 2.37 \\ 
  Max. & 2.00 & 2.97 & 1.55 & 11.89 & 5.17 & 86.60 \\ 
    \multicolumn{7}{c}{} \\
  \multicolumn{1}{c}{}  &   \multicolumn{6}{c}{Bivariate} \\
Min. & 1.01 & 0.59 & 0.68 & 0.70 & 0.60 & 0.55 \\ 
  1st Qu. & 1.34 & 0.90 & 1.04 & 1.10 & 0.92 & 0.84 \\ 
  Median & 1.45 & 1.07 & 1.17 & 1.28 & 1.03 & 0.99 \\ 
  3rd Qu. & 1.79 & 1.51 & 1.35 & 1.55 & 1.31 & 1.31 \\ 
  Max. & 2.29 & 188.67 & 2.88 & 21.02 & 16.58 & 76.41 \\ 
   \bottomrule
\end{tabular}
\label{relMSE_table_1_MCAR_30}
\end{table}

\begin{figure}[t!]
\centerline{\includegraphics[scale=.54]{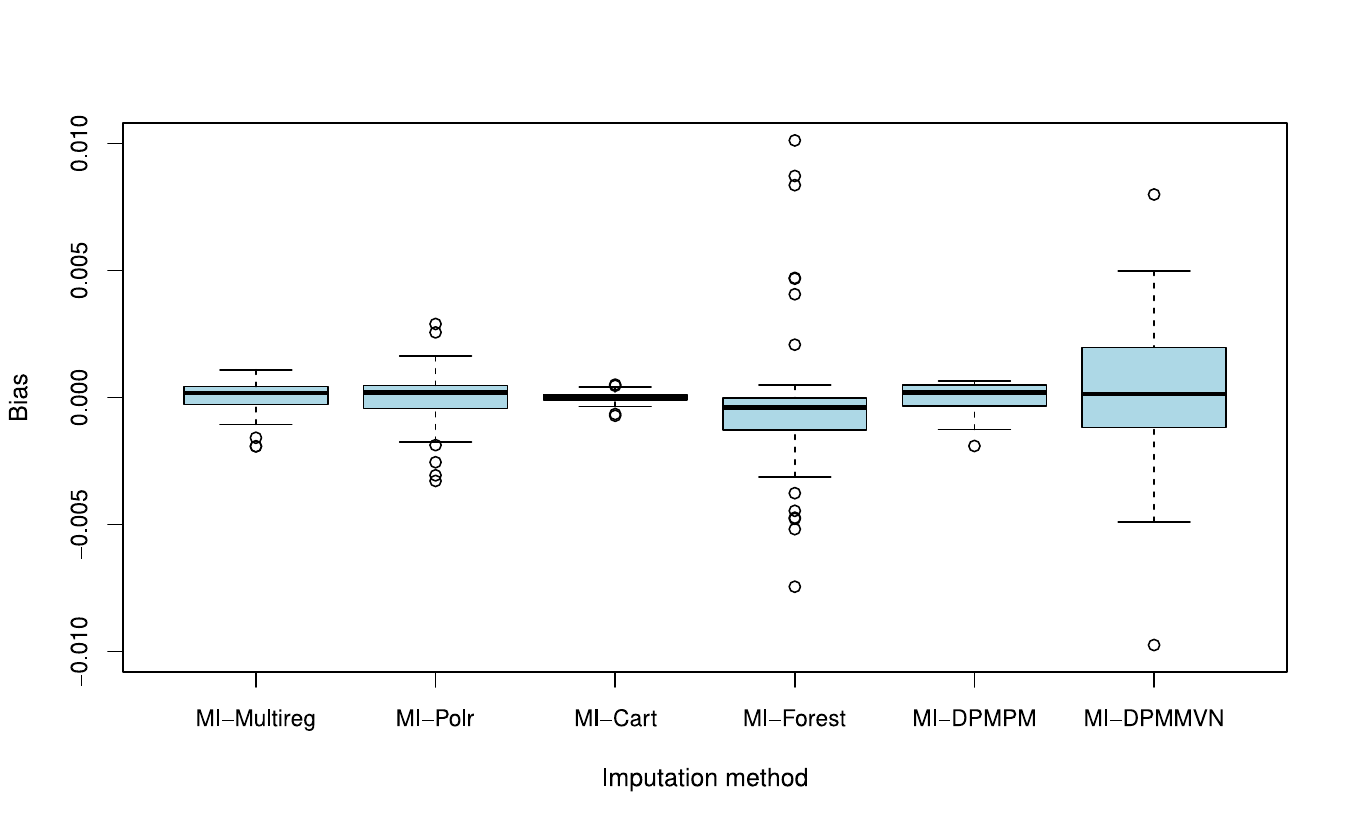}}
\caption{Distributions of estimated bias for marginal probabilities across all MI methods, under the 30\% MCAR scenario.}
\label{bias_plot_MCAR_30}
\end{figure}

Figure \ref{bias_plot_MCAR_30} displays the distributions of estimated bias for the marginal probabilities under each method. The estimated bias values are generally lower than those from the MAR scenarios.
The overall trends mimic the trends in the estimated relative MSE, at least for marginal estimands. MI-Cart again results in the least biased estimates for the marginal probabilities across all the methods. MI-Multireg, MI-Polr, and MI-DPMPM outperform MI-Forest and MI-DPMMVN. 
We present results for bivariate and trivariate probabilities, and numerical summaries of all the estimated bias values in the supplementary material.

As a final sensitivity analysis, we examine whether the overall conclusions from our first MCAR scenario change as the missingness rate increases, as we did under MAR. We do so by increasing the missingness rates to 45\% independently for the six variables.
\begin{table}[t!]
\centering
\footnotesize
\caption{Distributions of relative MSEs for marginal and bivariate probabilities across all MI methods, under the 45\% MCAR scenario.} 
\begin{tabular}{rrrrrrr}
  \toprule
 & MI-Multireg & MI-Polr & MI-Cart & MI-Forest & MI-DPMPM & MI-DPMMVN \\ 
  \midrule
    \multicolumn{1}{c}{}  &   \multicolumn{6}{c}{Marginal} \\
Min. & 1.57 & 1.43 & 1.45 & 1.62 & 1.44 & 0.74 \\ 
  1st Qu. & 1.76 & 1.78 & 1.71 & 2.02 & 1.75 & 1.36 \\ 
  Median & 1.91 & 1.98 & 1.77 & 2.62 & 1.90 & 2.62 \\ 
  3rd Qu. & 2.08 & 2.42 & 1.87 & 4.84 & 2.00 & 4.30 \\ 
  Max. & 4.59 & 4.96 & 2.10 & 28.64 & 11.57 & 198.28 \\ 
    \multicolumn{7}{c}{} \\
  \multicolumn{1}{c}{}  &   \multicolumn{6}{c}{Bivariate} \\
Min. & 1.09 & 0.47 & 0.60 & 0.65 & 0.59 & 0.46 \\ 
  1st Qu. & 1.69 & 0.94 & 1.12 & 1.25 & 0.95 & 0.86 \\ 
  Median & 1.89 & 1.27 & 1.34 & 1.62 & 1.17 & 1.17 \\ 
  3rd Qu. & 2.77 & 2.24 & 1.71 & 2.25 & 1.71 & 1.96 \\ 
  Max. & 4.62 & 413.60 & 7.29 & 63.35 & 38.58 & 173.94 \\ 
   \bottomrule
\end{tabular}
\label{relMSE_table_1_MCAR_45}
\end{table}
Table \ref{relMSE_table_1_MCAR_45} displays the distributions of the estimated relative MSE for the marginal and bivariate probabilities, under each method. As expected with a higher missingness rate, all relative MSE values are generally larger across all methods but the overall trends remain the same, with MI-Cart, and MI-DPMPM, once again yielding more accurate point estimates, although not by much, especially when focusing on the first to third quantiles of the distributions of relative MSEs.
We present additional results for this scenario in the supplementary material.

\section{Discussion} \label{conclusion}
Our simulation results suggest several general conclusions about the performances of the MI methods for ordinal data. Overall, MI-Cart and MI-DPMPM arguably outperform all other methods, resulting in coverage rates over 0.90 across all scenarios, the lowest relative MSE values, and the least biased estimates altogether. MI-Polr often yields comparable estimates especially when the missingness rate is 30\%, and is only slightly worse when the missingness rate increases to $45\%$. The performance of MI-Forest is often not too far behind the performance of MI-Polr. Since MI-Multireg requires more parameters compared to MI-Polr, it can yield slightly worse estimates when the data is missing under MAR, particularly as the missingness rate increases. While MI-DPMMVN generally yields less accurate estimates than MI-Cart, and MI-DPMPM, the estimates are not often far off particularly for the bivariate and trivariate estimands.

When handling missing values in ordinal data, clearly analysts are able to choose between treating the variables as either ordinal or nominal. Across most of the simulation scenarios, the results suggest that models developed specifically for ordinal data can provide very accurate estimates of marginal, bivariate, and even trivariate probabilities, comparable to MI-Cart and MI-DPMPM. However, there are also scenarios where models developed for ordinal data may not always provide clear advantages over the alternatives. Specifically, for GLM models for example, the results suggest that one might actually favor MI-Multireg over MI-Polr when dealing with MCAR scenarios with low missingness rates. Indeed, it is well known that the proportional odds model in MI-Polr may not always match the distributions in the data, so that the multinomial logistic models in MI-Multireg may be preferred in such cases. We recommend that analysts check the model assumptions carefully before selecting MI-Polr over MI-Multireg. Overall, when dealing with MI using GLM models, the results show that there can be advantages to treating the variable as nominal instead of ordinal, in certain scenarios. Similarly, MI-DPMPM generally outperforms MI-DPMMVN across all simulation scenarios, again suggesting that in the context of MI, there may not always be clear advantages for flexible ordinal models over other alternatives. However, when choosing between MI-DPMPM and MI-DPMMVN, we again suggest that analysts do extensive model checking to ascertain that the model assumptions are met.

Finally, as with any simulation study, the conclusions from our simulation scenarios may not generalize to other different settings. 
Also, we only consider data containing ordinal variables, whereas data from many other applications often include mixed data types. In those settings, it is possible that the trends we observe based on our results may change. Of all the methods we consider in our main study, only MI-Cart, MI-Forest and MI-DPMMVM can be applied directly to multiple data types (in the case of MI-DPMMVM, specifically to ordered categorical and continuous variables). 
Future work may focus on evaluating the performance of these methods in settings with mixed data types.

\section{Supplementary materials}
The supplementary material contains the data dictionary for the ACS dataset, the review of the MI combining rules, the logistic regressions for the MAR scenarios, introduction to GAIN and missForest as alternative MI methods, as well as additional tables and displays for all the simulation scenarios in Section \ref{result}, and histograms showing the marginal distributions of all variables in our population data from the ACS dataset.

\bibliographystyle{agsm}
\bibliography{references}

\end{document}